\documentclass[pra,showpacs,superscriptaddress,twocolumn]{revtex4}%
\usepackage{epsfig}
\usepackage{graphics}%
\usepackage{amsmath}%
\usepackage{amsfonts}%
\usepackage{amssymb}%
\usepackage{graphicx}
\begin{document}
\title{Ramsey-like measurement of the decoherence rate between Zeeman sub-levels}
\author{M. Shuker}
\affiliation{Department of Physics, Technion-Israel Institute of Technology, Haifa 32000, Israel}
\author{O. Firstenberg}
\affiliation{Department of Physics, Technion-Israel Institute of Technology, Haifa 32000, Israel}
\author{Y. Sagi}
\affiliation{Department of Physics of Complex Systems, Weizmann Institute of Science,
Rehovot 76100, Israel }
\author{A. Ben-kish}
\affiliation{Department of Physics, Technion-Israel Institute of Technology, Haifa 32000, Israel}
\author{N. Davidson}
\affiliation{Department of Physics of Complex Systems, Weizmann Institute of Science,
Rehovot 76100, Israel }
\author{A. Ron}
\affiliation{Department of Physics, Technion-Israel Institute of Technology, Haifa 32000, Israel}


\begin{abstract}
Two-photon processes that involve different sub-levels of the ground state of
an atom, are highly sensitive to depopulation and decoherence within the
ground state. For example, the spectral width of electromagnetically induced
transparency resonances in $\Lambda-$type system, are strongly affected by the
ground state depopulation and decoherence rates. We present a direct
measurement of decay rates between hyperfine and Zeeman sub-levels in the
ground state of $^{87}$Rb vapor. Similar to the \emph{relaxation-in-the-dark}
technique, pumping lasers are used to pre-align the atomic vapor in a well
defined quantum state. The free propagation of the atomic state is monitored
using a Ramsey-like method. Coherence times in the range $1-10$ ms were
measured for room temperature atomic vapor. In the range of the experimental
parameters used in this study, the dominant process inducing Zeeman
decoherence is the spin-exchange collisions between rubidium atoms.

\end{abstract}
\maketitle

\section{Introduction\label{sec_introduction}}

Decay processes within the ground state strongly affect the dynamics of
various two-photon processes. For example, in $\Lambda-$type
electromagnetically induced transparency (EIT) \cite{HarisEIT1997,Arimondo96},
two resonant laser fields couple two sub-levels within the ground state to a
common excited state. The decoherence rate between the ground-state
sub-levels, $\gamma_{12}$, determines the spectral width of the EIT, which in
turn affects all related phenomena. When the EIT medium is used to reduce the
group velocity of light pulses
\cite{HarrisSlowLight995,ScullyPRL1999UlraslowGroupVelocity,HauSlowing1999},
the decoherence rate influences the minimal group velocity achievable, and, in
storage of light experiments \cite{HauStorage2001,LukinPRL2001}, the
decoherence rate determines the possible storage duration. In these as well as
other applications of EIT (\textit{e.g.}, \cite{HarisPRL1999_NonLinear}) the
decoherence rate within the ground state is a key parameter that should be
carefully characterized.

Measurements of the decay rates within the ground state of various media has
been extensively studied. Several methods, such as Raman scattering
\cite{BookLaserSpectroscopy} and coherent anti-Stokes Raman scattering
\cite{BookNonLinearOptics} are commonly used for decay rate measurements.
Recently, an alternative method based on measuring the fluorescence from the
excited state during a coherent population trapping (CPT) experiment was
demonstrated \cite{ScullyPRA2007}. In that work, the atomic medium was driven
to a dark\ state using two lasers of orthogonal polarizations -- the pump and
the probe. A measurement of the fluorescence intensity from the excited level
was used to calculate the decoherence rate within the ground state.

Here, we propose and demonstrate a new method, analogous to Ramsey
spectroscopy, to measure the decoherence rate between Zeeman levels of the
ground state. We first drive the atomic ensemble to a well known dark-state
using pump and probe beams of equal intensity. Then, the driving beams are
turned off and an axial magnetic field is applied to induce oscillations
between the dark and the bright states of the medium. A series of weak light
pulses composed of the two driving beams allows us to probe this oscillation
in a non-destructive manner. By tracking the decay of the phase oscillations,
the decoherence rate is directly measured. We compared the measured
decoherence rate to detailed measurements we performed on the decay rate of
populations between hyperfine manifolds and between Zeeman sublevels. We find
that in our experiment, spin-exchange collisions are the dominant cause for
decoherence between Zeeman sub-levels.

In section \ref{sec_theoretical_model} we outline the theoretical model that
was used to analyze the results of the experiments described below. In section
\ref{sec_experimental_setup} we describe the experimental setup. The
experimental results are given in section \ref{sec_experimental_results}, and,
in section \ref{sec_discusstion_and_conclusions}, we discuss our results and conclude.

\section{Theoretical model\label{sec_theoretical_model}}

Our experiments are performed within the D1 transition of $^{87}$Rb. The
ground and upper states of this transition each consist of two hyperfine
manifolds, $F=1,2$, which in turn contain Zeeman manifolds, as depicted in
Fig. \ref{FigureExpSetup}.B. In the experiments described below, we tune the
laser to different transitions and use various polarizations. Due to the
complexity of the levels structure, a proper description of the experimental
results requires a model that takes into account the following elements:
resonant coupling, homogenous and non-homogenous broadening of the optical
transitions, influence of spectator levels, static magnetic field,
spin-exchange collisions, and non-spin-preserving decay in the ground-state.
Therefore, we have developed an elaborate numerical model, based on a program
package (toolbox) by S. M. Tan, which was specially designed for problems in
quantum optics \cite{QO_toolbox}.

The model describes a single $^{87}$Rb atom with all sixteen sub-levels in the
ground and first-excited states. It calculates the time-dependent solution of
the single-atom density matrix. The coupling with the laser fields is
calculated under the dipole and the rotating-wave approximations and accounts
for the various dipole moments and frequency detuning of all the allowed
transitions. The parameters that govern the one-photon transitions,
\textit{e.g.}, pressure broadening due to the buffer gas, were found from
independent measurements of the absorption spectrum for the specific vapor
cell used in the experiment. Doppler broadening is taken into account by
performing the simulation for different velocity groups and integrating over
the Boltzman distribution. Spin-exchange collisions are modeled by introducing
an additional mean-field atom, with which the system interacts, presenting the
density matrix of the two atoms in the product space of their electronic and
nuclear spins, and assuming decoherence between the singlet and triplet states
of the electronic part. This procedure is done in an iterative manner,
modifying the mean-field atom according to the state of the system with each
iteration. The model also include a simple linear decay mechanism which does
not preserve total spin and is responsible for the decay of the system to the
state of uniform population within the ground-state sub-levels (thermal
equilibrium). The experimental outcome, which is the absorption coefficient in
the present study, is calculated by taking the proper expectation value of the
time-dependent density-matrix.

\section{Experimental Setup\label{sec_experimental_setup}}

\begin{figure}
[ptb]
\begin{center}
\includegraphics[
height=2.9438in,
width=3.3918in
]%
{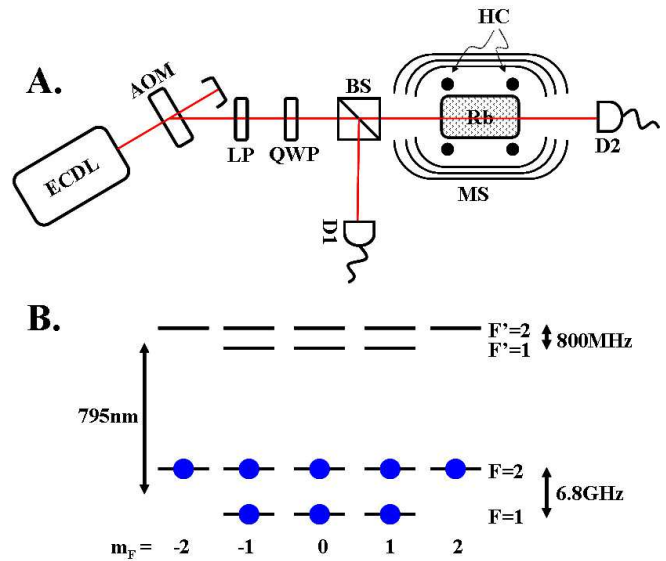}%
\caption{(color online) A. Experimental setup: ECDL - external cavity diode
laser ; AOM - acousto-optic modulator ; LP - linear polarizer ; QWP - quarter
wave-plate ; MS - magnetic shield ; HC - Helmholtz coils ; D1,D2 - detectors.
B. Energy levels scheme of D1 transition of $^{87}Rb$. Blue circles illustrate
the population at thermal equilibrium.}%
\label{FigureExpSetup}%
\end{center}
\end{figure}

The experimental setup is depicted in figure \ref{FigureExpSetup}.A. An
external cavity diode laser (ECDL) is locked to various transitions of $^{87}%
$Rb using a saturated absorption spectroscopy setup. Different hyperfine
transitions within the D1 transition of $^{87}$Rb ($\sim795$ nm) are used for
different measurements, as detailed below. The energy levels diagram of the D1
transition is depicted in figure \ref{FigureExpSetup}.B. The intensity of the
laser is controlled using an acousto-optic modulator (AOM). The first
diffraction order of the AOM is used for the experiment while the zero order
is sent to a beam stop. The polarization of the laser is set, either to linear
or to circular, by a linear polarizer and a quarter wave-plate. The laser beam
is sampled before entering the vapor cell using a beam-splitter and a
detector. The beam then passes through a $5$ cm long, $2.5$ cm diameter vapor
cell, and the transmitted intensity is measured on a second detector. By
comparing the readings of the two detectors the absorption coefficient of the
atomic medium is evaluated.%

The vapor cell contains isotopically pure $^{87}$Rb and $30$ Torr of Neon
buffer gas. The Rubidium vapor density in the cell is adjusted in the region
$1-9\times10^{11}$ $/cc$ by varying the temperature of the cell
\cite{SteckAlkaliData}. The temperature of the vapor cell is controlled using
non-magnetic heaters, and stabilized to within $\pm0.1$ $%
{{}^\circ}%
C$ (the heaters are switched off during the measurements). The beam waist
diameter inside the vapor cell is about $6$ mm. The vapor cell is placed
within a three-layered magnetic shield, which reduces the residual magnetic
field in the cell to less than $50$ $\mu$G. An axial magnetic field, $B_{z}$,
is applied in some of the measurements using a set of Helmholtz coils. This
applied magnetic field sets the quantization axis, $\widehat{z}$, to be
parallel to the propagation direction of the laser beams, $\widehat{k}$.

\section{Experimental Results\label{sec_experimental_results}}

Several measurements were performed to evaluate the different decay rates
within the ground state of the $^{87}$Rb atom. In all the measurements a
strong laser beam was used to pump the atoms to a specific quantum state. The
power of the laser beam during the pumping stage was $\sim1.5$ mW, and the
pumping duration was 2 seconds (much longer than required to reach
equilibrium). After the pumping stage the beam was turned off and the medium
relaxed to its thermal steady-state (\emph{relaxation in the dark}
\cite{FranzenPR1959}). The relaxation rates were measured by sending a series
of weak and short laser pulses ("probes") and monitoring their absorption. We
typically used a power of $10$ $\mu$W for the probes and their duty cycle was
about $5\%$ (so the average power during the probing stage was about $500$
nW). We verified that the probes were weak enough so their influence on the
populations and coherences of the medium was negligible. By tuning the pump
and the probes to various transitions and setting their polarization,
different decay rates within the ground state can be measured. We first
measured decay rates of populations as a reference for the main measurement of
the decoherence rate between Zeeman sublevels of the same hyperfine manifold.

\subsection{Decay of populations}

\begin{figure}
[ptb]
\begin{center}
\includegraphics[
height=2.5503in,
width=3.3918in
]%
{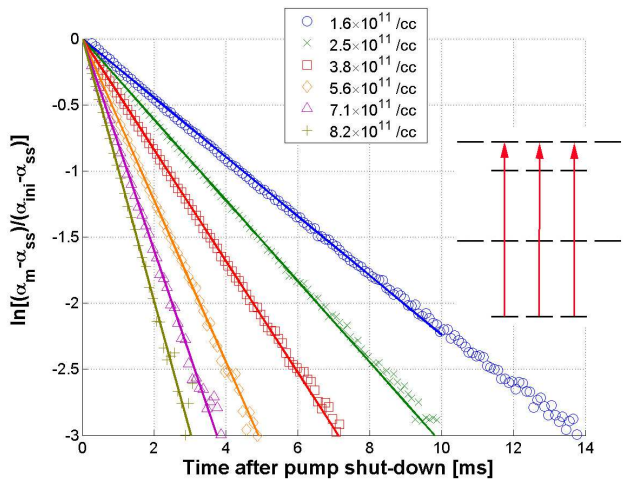}%
\caption{(color online) Return of hyperfine population to thermal steady state
after optical pumping, for different densities of the Rubidium vapor. Here
$\alpha_{m}$ is the measured absorption coefficient, $\alpha_{ss}$ is the
absorption coefficient at thermal steady state and $\alpha_{ini}$ is the
absorption coefficient at the end of the pumping stage. Each density shows a
linear line in a semilogarithmic graph, demonstrating that the decay is
exponential and the rate depends on the Rubidium density. The accuracy of
determining the decay rate (i.e. the slope) was better than 5\% for all the
densities we measured. The inset shows the transitions driven by the laser in
this measurement. }%
\label{FigureHyperfineDecay}%
\end{center}
\end{figure}

The laser is tuned to the $F=1\rightarrow F^{\prime}=2$ transition and set to
linear polarization. After the pumping stage most of the atoms populate the
$F=2$ level of the ground state manifold. After the pumping process reaches a
steady-state the pump beam is shut-down rapidly, and the rate of return of
population to the $F=1$ level of the ground-state is monitored by measuring
the absorption of the $F=1\rightarrow F^{\prime}=2$ transition. The fraction
of atoms populating the $F=1$ level is proportional to the absorption
coefficient, taking into account the various Clebsch-Gordan coefficients of
the different Zeeman levels. Fig. \ref{FigureHyperfineDecay} depicts the
measured absorption versus time, for different rubidium densities. It is
evident that the decay is exponential and that its rate depends on the
rubidium density. The measured decay rate is linear with the rubidium density
(see circles in Fig. \ref{FigureDecayRates}), showing that the dominant decay
mechanism of hyperfine population is Rb-Rb spin-exchange collisions. From
these measurements we calculated that the cross-section for Rb-Rb
spin-exchange collision is $\sigma_{Rb-Rb}=(2.05\pm0.2)\times10^{-14}cm^{2}$,
in good agreement with previous measurements \cite{HapperRMP1972}.%

\begin{figure}
[ptb]
\begin{center}
\includegraphics[
height=2.5227in,
width=3.3512in
]%
{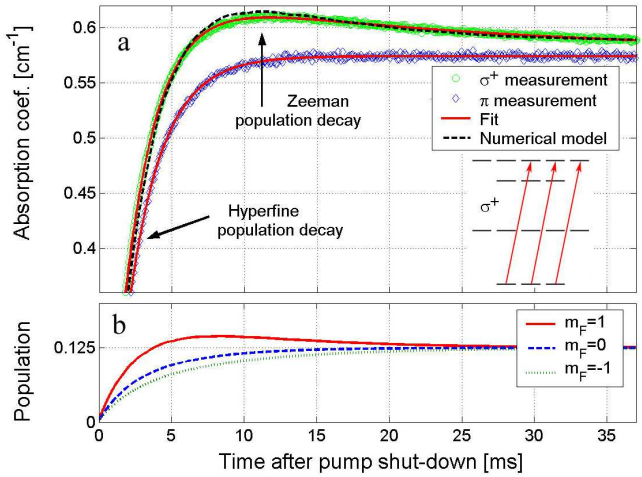}%
\caption{(color online) (a) Decay curves, after pumping with linear and
right-circular polarizations, on the F=1$\rightarrow$F'=2 transition. Both
measurements were performed at a vapor density of about $3.8\times10^{11}/cc$
(the linear polarization measurement is one those depicted in Fig.
\ref{FigureHyperfineDecay}). For the linear polarization (diamonds), a single
exponential decay is observed due to spin-exchange. For the circular
polarization (circles), a double exponential decay is observed: a fast decay
towards an absorption higher than the steady-state value and a slow decay to
the steady state absorption. We attribute the fast decay to hyperfine
population decay and the slow decay to Zeeman population decay. The red solid
lines are the respective exponential fits, while the black dashed line show
the result of our numerical model. The inset shows the transitions drived by
the laser in the measurement with circular polarization. (b) Results of the
numerical model, showing the populations of the three Zeeman sublevels of the
F=1 manifold, during the decay. At the end of the pumping stage, the three
level are nearly vacant, and they reach the equilibrium value of $1/8$ after
the decay. Evidently, the $m_{F}=1$ level is over-populated during the fast
decay, which results in an absorption higher than the steady-state value (due
to the different Clebsch-Gordan coefficients). }%
\label{FigureZeemanDecay}%
\end{center}
\end{figure}

In order to measure the decay rate of population difference between Zeeman
sub-levels, a pumping process which polarizes the medium was introduced. The
pump beam was tuned to the $F=1\rightarrow F^{\prime}=2$ transition and its
polarization was set to circularly positive ($\sigma^{+}$). A result of such a
measurement is depicted in Fig. \ref{FigureZeemanDecay}a alongside a reference
measurement with linear polarization (one of the measurements from Fig.
\ref{FigureHyperfineDecay}). While the linear polarization measurement shows a
single exponential decay, the circular polarization measurement shows an
unusual decay pattern. The population rapidly decays to \emph{higher}
absorption and then slowly decays down to the thermal steady-state absorption.
As in the linear polarization case, the pumping process empties the $F=1$
manifold. Since the pump is circularly polarized, the steady-state population
in the $F=2$ level is biased towards the higher $m_{F}$ states. After the pump
shut-off, spin-exchange collisions result in rapid transfer of population back
to the $F=1$ level. Since spin-exchange interaction conserves total angular
momentum of the atomic ensemble, the population returning to the $F=1$ level
is also biased towards higher $m_{F}$ states. Figure \ref{FigureZeemanDecay}b
shows the populations of the three $F=1$ Zeeman sub-levels versus time, as
calculated by our numerical model. When probing with the $F=1\rightarrow
F^{\prime}=2~;~\sigma^{+}$ transition, the $m_{F}$ population-bias towards
higher $m_{F}$ states results in higher absorption, since the Clebsch-Gordan
coefficients related to the $m_{F}=-1,0,+1$ states are higher for larger
$m_{F}$ ($\sqrt{1/12}$, $\sqrt{1/4}$, $\sqrt{1/2}$, respectively). The slow
decay of the absorption coefficient back to the steady-state absorption is the
Zeeman population decay rate (decay of polarization).

Both the fast (hyperfine) and the slow (Zeeman) decays fit well an exponential
decay, and hence the experimental data can be fitted well by a double
exponential decay or by a single exponential decay in the case of linear
polarization. The fast decay rate fits, within the experimental error, the
results obtained in the linear polarization measurement, depicted in Fig.
\ref{FigureHyperfineDecay}. The measured slow decay rate is $\sim50$ $s^{-1}$,
in a good agreement with the theoretical prediction of the diffusion induced
decay \cite{PappasPRL1981_diffusion_decay_rate}, taking into account both the
laser beam diameter and the vapor cell diameter. Therefore, the Zeeman
population decay rate is nearly constant at different rubidium densities, as
depicted in Fig. \ref{FigureDecayRates} (squares), and the small slope is
partially due to the increased diffusion at the higher temperatures.  We
conclude that, in our setup, the dominant mechanism for Zeeman population
decay is the diffusion of the atoms in and out of the laser beam and wall-collisions.

In other pump-probe configurations \cite{FranzenPR1959} it is possible to
obtain a decay curve in which both the hyperfine and the Zeeman decays are in
the same direction. For example, if the pump and probe are tuned to
the\ $F=2\rightarrow F^{\prime}=1~;~\sigma^{+}$ transition, both decays
increase the absorption coefficient of the medium. The configuration presented
in Fig. \ref{FigureZeemanDecay} has the benefit that the fast and slow decays
are at opposite directions, making the data analysis easier. We have measured
the Zeeman population decay rate in both configurations and obtained similar results.%

\subsection{Decay of coherence between Zeeman sub-levels}

\begin{figure}
[ptb]
\begin{center}
\includegraphics[
height=2.5227in,
width=3.3512in
]%
{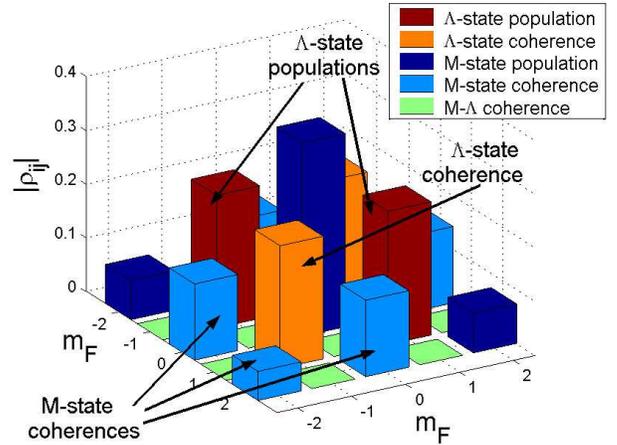}%
\caption{(color online) Theoretical calculation, using our numerical model, of
the reduced density matrix for the $F=2$ ground-state manifold at the end of
the pumping stage (steady-state solution). Two mixed dark states arise: the
$\Lambda-$state, which involves sub-levels $m_{F}=-1,+1$, and the $M-$state,
which involves sub-levels $m_{F}=-2,0,+2$. The two dark-states are equally
populated.}%
\label{fig_steady_state}%
\end{center}
\end{figure}

\begin{figure}
[ptb]
\begin{center}
\includegraphics[
height=2.5227in,
width=3.3512in
]%
{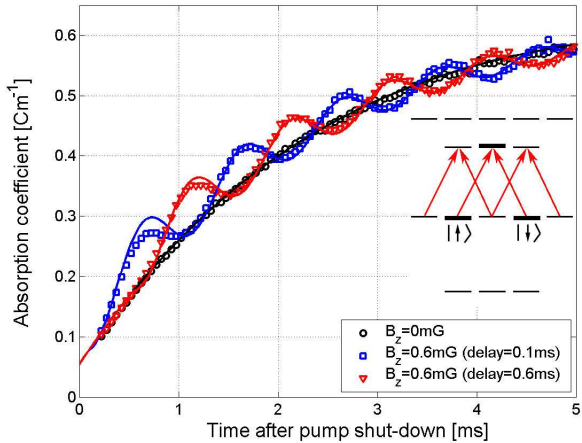}%
\caption{(color online) Decay of the induced transparency and dark-to-bright
state oscillations caused by an axial magnetic field. When no magnetic field
is applied (black circles), a regular decay back to thermal absorption is
observed. When an axial magnetic field is applied $0.1$ ms after the pump
shut-down (blue squares) oscillations in the absorption are observed. By
applying the magnetic field after a longer delay (red triangles), the same
oscillations appear at opposite phase - allowing us to extract the
oscillations component of the signal with high accuracy. Solid lines show the
results of our numerical model for the three cases, demonstrating a good
agreement with the experimental results. An animation of the reduced density
matrix for the $F=2 $ manifold (as depicted in Fig. \ref{fig_steady_state}),
during the decoherence measurement is given in \cite{EPAPS_Remark}. The inset
shows the transitions drived by the lasers in this measurement.}%
\label{FigureDarkToBright}%
\end{center}
\end{figure}

In this section we demonstrate a simple technique, analogous to Ramsey
spectroscopy, which directly measures the decoherence rate between Zeeman
sub-levels of the $F=2$ hyperfine manifold of the $^{87}$Rb ground state. The
first step of the measurement is to create a \emph{coherent superposition} of
the relevant Zeeman sub-levels. For that purpose the medium is driven on the
$F=2\rightarrow F^{\prime}=1$ transition with $\sigma^{+}$and $\sigma^{-}$
polarizations with equal intensities and a well defined relative phase. The
atomic medium is optically pumped until it reaches a steady state. The quantum
state of the atomic system under the influence of the pumping lasers was
calculated using a semi-classical model considering all sixteen relevant
energy levels. The steady state solution is comprised of an incoherent mixture
of two dark-states in the $F=2$ manifold of the ground state. The, so-called,
$\Lambda$-state,%
\[
\left\vert \Lambda\right\rangle =\frac{1}{\sqrt{2}}\left(  \left\vert
m_{F}=-1\right\rangle +\left\vert m_{F}=+1\right\rangle \right)  ,
\]
and the, so-called, $M-$state,%
\[
\left\vert M\right\rangle =\frac{1}{\sqrt{8}}\left(  \left\vert m_{F}%
=-2\right\rangle +\sqrt{6}\cdot\left\vert m_{F}=0\right\rangle +\left\vert
m_{F}=+2\right\rangle \right)  .
\]
The three levels involved in the formation of the $\Lambda$-state are marked
by thick lines in the inset of Fig. \ref{FigureDarkToBright}. These two
dark-states are evident in Fig. \ref{fig_steady_state} which depicts the
reduced density matrix of the five Zeeman sub-levels in the $F=2$ manifold.
Note that in the $\Lambda-$state both sub-levels have equal populations, while
in the $M-$state the $m_{F}=0$ sub-level holds most of the population. This
difference is due to the Clebsch-Gordan coefficients of the different optical
transitions. Note that a large fraction of the atomic population accumulates
in the various Zeeman sub-levels of the $F=1$ manifold (not shown in Fig.
\ref{fig_steady_state}).%

After the steady state was achieved, the pumping lasers are turned off and a
small axial magnetic field, $B_{Z}\simeq1$mG, is applied to the atomic medium.
The magnetic field slightly shifts the energies of the Zeeman sub-levels and
induces oscillations in their relative phases, at the Larmor frequency. For
the special case of a dark-state with equal populations in its two lower
levels (the $\Lambda$-state in our experiment), it can be easily converted to
a bright state by changing the relative phase between the lower levels. The
said bright-state is given by%
\[
\left\vert \Lambda^{\ast}\right\rangle =\frac{1}{\sqrt{2}}\left(  \left\vert
m_{F}=-1\right\rangle -\left\vert m_{F}=+1\right\rangle \right)  .
\]
Hence, oscillations of the phase between the lower levels (induced by the
applied axial magnetic field) results in oscillations between the dark the
bright states ($\left\vert \Lambda^{\ast}\right\rangle $ and $\left\vert
\Lambda\right\rangle $ respectively). These 'dark-to-bright' oscillations can
be detected by measuring changes in the transparency of the medium to probes
composed of $\sigma^{+}$ and $\sigma^{-}$ with equal intensity (to which the
medium is dark or bright). Decoherence between the lower levels will result in
the decay of these oscillations, so decay in the amplitude of the oscillations
is a direct measure of the decoherence between Zeeman sub-levels of a single
hyperfine manifold. This relation was also verified by our numerical model.

For the case of a dark-state with non-equal populations in the lower levels
($M-$state in our experiment), similar oscillations occur but with lower
visibility. The amplitude of the oscillations originating from the $M-$state
was calculated using our theoretical model, and should contribute about half
of the total oscillation. It is interesting to note that the effect of the
axial magnetic field on the $M-$state results also in very weak oscillations
with twice the Larmor frequency. These oscillations are not observable in our
experimental system due to their low amplitude.%

\begin{figure}
[ptb]
\begin{center}
\includegraphics[
height=2.5503in,
width=3.3918in
]%
{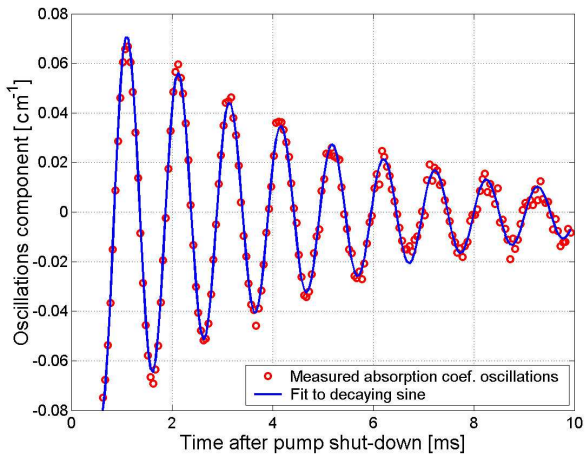}%
\caption{(color online) The signal of dark-to-bright state oscillations, which
was extracted from the data in Fig. \ref{FigureDarkToBright}. A decaying
sinusoidal function fits the measured data well. }%
\label{FigureDarkToBrightOscOnly}%
\end{center}
\end{figure}

Fig. \ref{FigureDarkToBright} depicts the results of a Zeeman decoherence
measurement. When no magnetic field is applied ($B_{Z}=0$), a decay of the
transparency is observed (blue circles), which occurs due to decoherence as
well as other decay mechanisms. When a small, DC, magnetic field is applied
($B_{Z}\neq0$) after the pump is shut down, oscillations in the transparency
are clearly apparent -- a result of the medium's oscillations from the dark to
the bright state and vice versa (green squares). Due to technical
considerations, the magnetic field is switched on with a delay of 0.1 ms after
the pump is shut-down. The cycle time of the observed oscillations is
determined by the magnitude of the applied magnetic field, and their amplitude
depends mostly on the magnitude of the coherence terms between Zeeman
sub-levels. Immediately after the pump is shut-down, the medium is at the dark
state. When the magnetic field is switched on, an increase in the absorption
coefficient is observed, followed by a decrease back to the $B_{Z}=0$ curve.
After a few cycles, the minima of the oscillations decrease below the
$B_{Z}=0$ curve, due to a small residual magnetic field which causes slow
oscillations even when no magnetic field is applied. In order to extract, with
high accuracy, the oscillation component of the signal, a second measurement
was performed with a longer delay between the pump shut-down and the
application of the magnetic field. The delay was chosen to be about 0.6 ms
(red triangles), so that the two measurements will have a phase-shift of $\pi
$. By subtracting the two measurements, the oscillations component can be
extracted from the signal, as depicted in Fig. \ref{FigureDarkToBrightOscOnly}%
. The oscillation component is well fitted by a sinusoidal function multiplied
by an exponential decay whose rate is equivalent to the decoherence rate
between the Zeeman levels of the $F=2$ hyperfine manifold. We have repeated
this measurement for different rubidium densities, and found that the decay
rate is similar to the hyperfine population decay rate (see Fig.
\ref{FigureDecayRates} and section \ref{sec_discusstion_and_conclusions}). By
repeating the measurements with different applied magnetic filed, we have
verified that the measured decoherence rate is independent of the applied
field's magnitude. %

The decoherence measurement presented here is analogues to the Ramsey
spectroscopy method. For the purpose of illustration we consider the
$\Lambda-$state, and denote its two lower levels $\left\vert \uparrow
\right\rangle =\left\vert F=2;m_{F}=-1\right\rangle $ and $\left\vert
\downarrow\right\rangle =\left\vert F=2;m_{F}=+1\right\rangle $ as depicted in
the inset of Fig. \ref{FigureDarkToBright}. As explained above, the pumping
process creates a coherent super-position of the two levels, given by
$1/\sqrt{2}\left(  \left\vert \uparrow\right\rangle +\left\vert \downarrow
\right\rangle \right)  $. After the pumping fields are turned off, the
super-position's phase is allowed to oscillate. In the present experiment,
since the two levels $\left\vert \uparrow\right\rangle ,\left\vert
\downarrow\right\rangle $ are degenerate, a small magnetic field is applied to
drive phase oscillations. After a certain time delay, the phase between the
two levels is detected by sending two laser pulses to which the transparency
depends on the phase of the super-position. By repeating this measurement for
different time delays, a graph monitoring the phase oscillations, as well as
the loss of coherence is obtained.

\section{Discussion and conclusions\label{sec_discusstion_and_conclusions}}

\begin{figure}
[ptb]
\begin{center}
\includegraphics[
height=2.5503in,
width=3.3918in
]%
{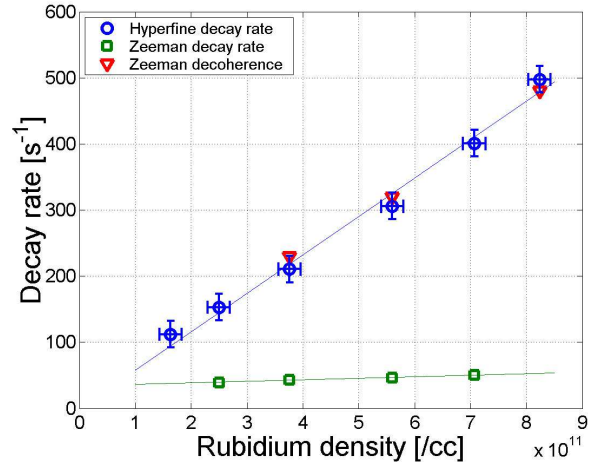}%
\caption{(color online) Measured decay rates vs. rubidium atomic density:
hyperfine population decay (blue circles), Zeeman population decay (green
squares) and Zeeman decoherence (red triangles). The solid lines are a guide
to the eye.}%
\label{FigureDecayRates}%
\end{center}
\end{figure}

We have presented measurements of various decay rates in the ground state
manifold of atomic rubidium. The experimental technique is based on the
relaxation-in-the-dark method \cite{FranzenPR1959}. By tuning the laser to
different transitions and polarizations, as well as switching on a magnetic
field at the probing stage, we were able to measure independently the decay
rates of population between hyperfine and Zeeman levels, and the decoherence
rate between Zeeman sub-levels of the same hyperfine manifold. The results of
all the measurements are presented in Fig. \ref{FigureDecayRates}.%

The decay of populations between hyperfine levels is dominated by
spin-exchange collisions between rubidium atoms, and hence depends linearly on
the rubidium density. The decay of population between Zeeman sub-levels,
\textit{i.e.}, the decay of polarization, is dominated by the diffusion of
atoms in the vapor cell, and is nearly independent of the rubidium density.
The decoherence rate between Zeeman sub-levels of the same hyperfine manifold
is equal to the hyperfine population decay rate within the measurement
accuracy. We conclude that no additional decay mechanism contributes to the
decoherence rate without affecting the decay of populations. Unlike optical
transitions, where the decoherence rate can be half the population decay rate
\cite{AtomPhotonInteractions}, we find that the decoherence rate between
ground-state levels is similar to the population decay rate. We attribute this
to the fact that both levels are affected by the population decay.

The various decay rates within the ground state are the key parameters of any
two-photon interaction, \textit{e.g.}, in EIT. The decay rates directly affect
all the features of the EIT phenomena, such as slow light, storage of light,
frequency standards, etc. Simple techniques to measure the decoherence rate
are important for many practical purposes and may be useful at different
realizations of two-photon phenomena. In the current experiment we measured
the decoherence rate between Zeeman sub-levels of the same hyperfine manifold.
The same method can be applied for measuring the decoherence rate between
levels belonging to different hyperfine manifolds. In addition, this method
can be adapted to any medium exhibiting EIT, by applying an analogous
technique to oscillate the relative phase between the lower levels.

\begin{acknowledgments}
We wish to acknowledge the help of Amnon Fisher, Yoav Erlich and Igal Levi.
This research was partially supported by the fund for encouragement of
research in the Technion.
\end{acknowledgments}

\bibliographystyle{apsrev}
\bibliography{references_decoherence}

\begin{thebibliography}{18}
\expandafter\ifx\csname natexlab\endcsname\relax\def\natexlab#1{#1}\fi
\expandafter\ifx\csname bibnamefont\endcsname\relax
  \def\bibnamefont#1{#1}\fi
\expandafter\ifx\csname bibfnamefont\endcsname\relax
  \def\bibfnamefont#1{#1}\fi
\expandafter\ifx\csname citenamefont\endcsname\relax
  \def\citenamefont#1{#1}\fi
\expandafter\ifx\csname url\endcsname\relax
  \def\url#1{\texttt{#1}}\fi
\expandafter\ifx\csname urlprefix\endcsname\relax\def\urlprefix{URL }\fi
\providecommand{\bibinfo}[2]{#2}
\providecommand{\eprint}[2][]{\url{#2}}

\bibitem[{\citenamefont{Harris}(1997)}]{HarisEIT1997}
\bibinfo{author}{\bibfnamefont{S.~E.} \bibnamefont{Harris}},
  \bibinfo{journal}{Physics Today} \textbf{\bibinfo{volume}{50}},
  \bibinfo{pages}{36} (\bibinfo{year}{1997}).

\bibitem[{\citenamefont{Arimondo}(1996)}]{Arimondo96}
\bibinfo{author}{\bibfnamefont{E.}~\bibnamefont{Arimondo}},
  \emph{\bibinfo{title}{Progress in Optics}}, vol.~\bibinfo{volume}{35}
  (\bibinfo{publisher}{Elsevier}, \bibinfo{address}{Amsterdam},
  \bibinfo{year}{1996}).

\bibitem[{\citenamefont{Kasapi et~al.}(1995)\citenamefont{Kasapi, Jain, Yin,
  and Harris}}]{HarrisSlowLight995}
\bibinfo{author}{\bibfnamefont{A.}~\bibnamefont{Kasapi}},
  \bibinfo{author}{\bibfnamefont{M.}~\bibnamefont{Jain}},
  \bibinfo{author}{\bibfnamefont{G.~Y.} \bibnamefont{Yin}}, \bibnamefont{and}
  \bibinfo{author}{\bibfnamefont{S.~E.} \bibnamefont{Harris}},
  \bibinfo{journal}{Phys. Rev. Lett.} \textbf{\bibinfo{volume}{74}},
  \bibinfo{pages}{2447} (\bibinfo{year}{1995}).

\bibitem[{\citenamefont{Kash et~al.}(1999)\citenamefont{Kash, Sautenkov,
  Zibrov, Hollberg, Welch, Lukin, Rostovtsev, Fry, and
  Scully}}]{ScullyPRL1999UlraslowGroupVelocity}
\bibinfo{author}{\bibfnamefont{M.~M.} \bibnamefont{Kash}},
  \bibinfo{author}{\bibfnamefont{V.~A.} \bibnamefont{Sautenkov}},
  \bibinfo{author}{\bibfnamefont{A.~S.} \bibnamefont{Zibrov}},
  \bibinfo{author}{\bibfnamefont{L.}~\bibnamefont{Hollberg}},
  \bibinfo{author}{\bibfnamefont{G.~R.} \bibnamefont{Welch}},
  \bibinfo{author}{\bibfnamefont{M.~D.} \bibnamefont{Lukin}},
  \bibinfo{author}{\bibfnamefont{Y.}~\bibnamefont{Rostovtsev}},
  \bibinfo{author}{\bibfnamefont{E.~S.} \bibnamefont{Fry}}, \bibnamefont{and}
  \bibinfo{author}{\bibfnamefont{M.~O.} \bibnamefont{Scully}},
  \bibinfo{journal}{Phys. Rev. Lett.} \textbf{\bibinfo{volume}{82}},
  \bibinfo{pages}{5229} (\bibinfo{year}{1999}).

\bibitem[{\citenamefont{Hau et~al.}(1999)\citenamefont{Hau, Harris, Dutton, and
  Behroozi}}]{HauSlowing1999}
\bibinfo{author}{\bibfnamefont{L.~V.} \bibnamefont{Hau}},
  \bibinfo{author}{\bibfnamefont{S.~E.} \bibnamefont{Harris}},
  \bibinfo{author}{\bibfnamefont{Z.}~\bibnamefont{Dutton}}, \bibnamefont{and}
  \bibinfo{author}{\bibfnamefont{C.~H.} \bibnamefont{Behroozi}},
  \bibinfo{journal}{Nature} \textbf{\bibinfo{volume}{397}},
  \bibinfo{pages}{594} (\bibinfo{year}{1999}).

\bibitem[{\citenamefont{Liu et~al.}(2001)\citenamefont{Liu, Dutton, Behroozi,
  and Hau}}]{HauStorage2001}
\bibinfo{author}{\bibfnamefont{C.}~\bibnamefont{Liu}},
  \bibinfo{author}{\bibfnamefont{Z.}~\bibnamefont{Dutton}},
  \bibinfo{author}{\bibfnamefont{C.~H.} \bibnamefont{Behroozi}},
  \bibnamefont{and} \bibinfo{author}{\bibfnamefont{L.~V.} \bibnamefont{Hau}},
  \bibinfo{journal}{Nature} \textbf{\bibinfo{volume}{409}},
  \bibinfo{pages}{490} (\bibinfo{year}{2001}).

\bibitem[{\citenamefont{Phillips et~al.}(2001)\citenamefont{Phillips,
  Fleischhauer, Mair, Walsworth, and Lukin}}]{LukinPRL2001}
\bibinfo{author}{\bibfnamefont{D.~F.} \bibnamefont{Phillips}},
  \bibinfo{author}{\bibfnamefont{A.}~\bibnamefont{Fleischhauer}},
  \bibinfo{author}{\bibfnamefont{A.}~\bibnamefont{Mair}},
  \bibinfo{author}{\bibfnamefont{R.~L.} \bibnamefont{Walsworth}},
  \bibnamefont{and} \bibinfo{author}{\bibfnamefont{M.~D.} \bibnamefont{Lukin}},
  \bibinfo{journal}{Phys. Rev. Lett.} \textbf{\bibinfo{volume}{86}},
  \bibinfo{pages}{783} (\bibinfo{year}{2001}).

\bibitem[{\citenamefont{Harris and Hau}(1999)}]{HarisPRL1999_NonLinear}
\bibinfo{author}{\bibfnamefont{S.~E.} \bibnamefont{Harris}} \bibnamefont{and}
  \bibinfo{author}{\bibfnamefont{L.~V.} \bibnamefont{Hau}},
  \bibinfo{journal}{Phys. Rev. Lett.} \textbf{\bibinfo{volume}{82}},
  \bibinfo{pages}{4611} (\bibinfo{year}{1999}).

\bibitem[{\citenamefont{Demtröder}(2002)}]{BookLaserSpectroscopy}
\bibinfo{author}{\bibfnamefont{W.}~\bibnamefont{Demtröder}},
  \emph{\bibinfo{title}{Laser Spectroscopy}}
  (\bibinfo{publisher}{Springer-Verlag}, \bibinfo{address}{New York},
  \bibinfo{year}{2002}).

\bibitem[{\citenamefont{Shen}(1984)}]{BookNonLinearOptics}
\bibinfo{author}{\bibfnamefont{Y.~R.} \bibnamefont{Shen}},
  \emph{\bibinfo{title}{Principles of Non-linear Optic}}
  (\bibinfo{publisher}{Wiley}, \bibinfo{address}{New York},
  \bibinfo{year}{1984}).

\bibitem[{\citenamefont{Patnaik et~al.}(2007)\citenamefont{Patnaik, Hsu,
  Agarwal, Welch, and Scully}}]{ScullyPRA2007}
\bibinfo{author}{\bibfnamefont{A.~K.} \bibnamefont{Patnaik}},
  \bibinfo{author}{\bibfnamefont{P.~S.} \bibnamefont{Hsu}},
  \bibinfo{author}{\bibfnamefont{G.~S.} \bibnamefont{Agarwal}},
  \bibinfo{author}{\bibfnamefont{G.~R.} \bibnamefont{Welch}}, \bibnamefont{and}
  \bibinfo{author}{\bibfnamefont{M.~O.} \bibnamefont{Scully}},
  \bibinfo{journal}{Phys. Rev. A} \textbf{\bibinfo{volume}{75}},
  \bibinfo{eid}{023807} (\bibinfo{year}{2007}).

\bibitem[{\citenamefont{Tan}(1999)}]{QO_toolbox}
\bibinfo{author}{\bibfnamefont{S.~M.} \bibnamefont{Tan}},
  \bibinfo{journal}{Journal of Optics B: Quantum and Semiclassical Optics}
  \textbf{\bibinfo{volume}{1}}, \bibinfo{pages}{424} (\bibinfo{year}{1999}).

\bibitem[{\citenamefont{Steck}(2003)}]{SteckAlkaliData}
\bibinfo{author}{\bibfnamefont{D.~A.} \bibnamefont{Steck}},
  \bibinfo{type}{Tech. Rep.} \bibinfo{number}{LA-UR-03-8638},
  \bibinfo{institution}{LANL} (\bibinfo{year}{2003}).

\bibitem[{\citenamefont{Franzen}(1959)}]{FranzenPR1959}
\bibinfo{author}{\bibfnamefont{W.}~\bibnamefont{Franzen}},
  \bibinfo{journal}{Phys. Rev.} \textbf{\bibinfo{volume}{115}},
  \bibinfo{pages}{850} (\bibinfo{year}{1959}).

\bibitem[{\citenamefont{Happer}(1972)}]{HapperRMP1972}
\bibinfo{author}{\bibfnamefont{W.}~\bibnamefont{Happer}},
  \bibinfo{journal}{Rev. Mod. Phys.} \textbf{\bibinfo{volume}{44}},
  \bibinfo{pages}{169} (\bibinfo{year}{1972}).

\bibitem[{\citenamefont{Pappas et~al.}(1981)\citenamefont{Pappas, Forber,
  Quivers, Dasari, Feld, and Murnick}}]{PappasPRL1981_diffusion_decay_rate}
\bibinfo{author}{\bibfnamefont{P.~G.} \bibnamefont{Pappas}},
  \bibinfo{author}{\bibfnamefont{R.~A.} \bibnamefont{Forber}},
  \bibinfo{author}{\bibfnamefont{W.~W.} \bibnamefont{Quivers}},
  \bibinfo{author}{\bibfnamefont{R.~R.} \bibnamefont{Dasari}},
  \bibinfo{author}{\bibfnamefont{M.~S.} \bibnamefont{Feld}}, \bibnamefont{and}
  \bibinfo{author}{\bibfnamefont{D.~E.} \bibnamefont{Murnick}},
  \bibinfo{journal}{Phys. Rev. Lett.} \textbf{\bibinfo{volume}{47}},
  \bibinfo{pages}{236} (\bibinfo{year}{1981}).

\bibitem[{EPA()}]{EPAPS_Remark}
\bibinfo{note}{See EPAPS Document No. ??? for a movie showing the dynamics of
  the reduced density matrix during the Zeeman decoherence measurement. For
  more information on EPAPS, see http://www.aip.org/pubservs/epaps.html.}

\bibitem[{\citenamefont{Cohen-Tannoudji
  et~al.}(1998)\citenamefont{Cohen-Tannoudji, Dupont-Roc, and
  Grynberg}}]{AtomPhotonInteractions}
\bibinfo{author}{\bibfnamefont{C.}~\bibnamefont{Cohen-Tannoudji}},
  \bibinfo{author}{\bibfnamefont{J.}~\bibnamefont{Dupont-Roc}},
  \bibnamefont{and} \bibinfo{author}{\bibfnamefont{G.}~\bibnamefont{Grynberg}},
  \emph{\bibinfo{title}{Atom-Photon Interactions : Basic Processes and
  Applications}} (\bibinfo{publisher}{Wiley-Interscience},
  \bibinfo{year}{1998}).

\end{thebibliography}

\end{document}